\documentclass[conference]{IEEEtran}
\usepackage{cite}
\usepackage{amsmath,amssymb,amsfonts}
\usepackage{dblfloatfix}    
\usepackage{graphicx}
\usepackage{textcomp}
\usepackage{xcolor}
\usepackage{fancyhdr}
\usepackage[hyphens]{url}

\usepackage[T1]{fontenc}
\usepackage{algorithm}
\usepackage{algpseudocode}
\usepackage{color}
\usepackage{lipsum}
\usepackage{float}
\usepackage{setspace}
\usepackage{subcaption}
\usepackage{caption}
\usepackage{bm}
\usepackage{eso-pic}
\usepackage{listings}

\definecolor{darkWhite}{rgb}{0.94,0.94,0.94}
\definecolor{codegreen}{rgb}{0,0.6,0}
\definecolor{codegray}{rgb}{0.5,0.5,0.5}
\definecolor{codepurple}{rgb}{0.58,0,0.82}
\definecolor{backcolour}{rgb}{0.95,0.95,0.92}
\lstdefinestyle{mystyle}{
    backgroundcolor=\color{backcolour},
    commentstyle=\color{codegreen},
    keywordstyle=\color{magenta},
    numberstyle=\tiny\color{codegray},
    stringstyle=\color{codepurple},
    basicstyle=\footnotesize,
    breakatwhitespace=false,
    breaklines=true,
    captionpos=b,
    keepspaces=true,
    numbers=left,
    numbersep=5pt,
    showspaces=false,
    showstringspaces=false,
    showtabs=false,
    tabsize=2
}

\def\BibTeX{{\rm B\kern-.05em{\sc i\kern-.025em b}\kern-.08em
    T\kern-.1667em\lower.7ex\hbox{E}\kern-.125emX}}

\usepackage{tikz}
\usetikzlibrary{decorations.pathreplacing,calc}

\newcommand{\placetextbox}[3]{
  \setbox0=\hbox{#3}
  \AddToShipoutPictureFG*{
    \put(\LenToUnit{#1\paperwidth},\LenToUnit{#2\paperheight}){\vtop{{\null}\makebox[0pt][c]{#3}}}%
  }%
}%

\algdef{SE}[SUBALG]{Indent}{EndIndent}{}{\algorithmicend\ }%
\algtext*{Indent}
\algtext*{EndIndent}

\newcommand*\mycircle[1]{
	\raisebox{.5pt}{\textcircled{\raisebox{-.9pt} {#1}}}
}

\fancypagestyle{firstpage}{
  \fancyhf{}
  
  \fancyhead[C]{\vspace{10pt}\normalsize{FPL 2023 Submission ID
      \textbf{\#2671}}\\\vspace{-25pt}}
  \fancyfoot[C]{\thepage}
}

\title{An Open-Source Framework for Efficient Numerically-Tailored Computations}

\author{\IEEEauthorblockN{Louis Ledoux}
\IEEEauthorblockA{
\textit{Barcelona Supercomputing Center (BSC)}\\
\textit{Universitat Politecnica de Catalunya (UPC)}\\
louis.ledoux@bsc.es}
\and
\IEEEauthorblockN{Marc Casas}
\IEEEauthorblockA{
\textit{Barcelona Supercomputing Center (BSC)}\\
\textit{Universitat Politecnica de Catalunya (UPC)}\\
marc.casas@bsc.es}}

\begin{document}
\IEEEoverridecommandlockouts
\IEEEaftertitletext{\vspace{-2\baselineskip}}
\maketitle
\thispagestyle{firstpage}
\pagestyle{plain}


\begin{abstract}
We present a versatile open-source framework designed to facilitate efficient, numerically-tailored Matrix-Matrix Multiplications (MMMs).
The framework offers two primary contributions: first, a fine-tuned, automated pipeline for arithmetic datapath generation, enabling highly customizable systolic MMM kernels; second, seamless integration of the generated kernels into user code, irrespective of the programming language employed, without necessitating modifications.

We employ this framework within a cutting-edge platform, comprising a Power9 host, an OpenCAPI link, and a Xilinx Virtex UltraScale+ FPGA.
The framework demonstrates a systematic enhancement in accuracy per energy cost across diverse High Performance Computing (HPC) workloads displaying a variety of numerical requirements, such as Artificial Intelligence (AI) inference and Sea Surface Height (SSH) computation.
For AI inference, we consider a set of state-of-the-art neural network models, namely ResNet18, ResNet34, ResNet50, DenseNet121, DenseNet161, DenseNet169, and VGG11, in conjunction with two datasets, two computer formats, and 27 distinct intermediate arithmetic datapaths.
Our approach consistently reduces energy consumption across all cases, with a notable example being the reduction by factors of $3.3\times$ for IEEE754-32 and $1.4\times$ for Bfloat16 during ImageNet inference with ResNet50.
This is accomplished while maintaining accuracies of $82.3\%$ and $86\%$, comparable to those achieved with conventional Floating-Point Units (FPUs).
In the context of SSH computation, our method achieves fully-reproducible results using double-precision words, surpassing the accuracy of conventional double- and quad-precision arithmetic in FPUs.
Our approach enhances SSH computation accuracy by a minimum of $5\times$ and $27\times$ compared to IEEE754-64 and IEEE754-128, respectively, resulting in $5.6\times$ and $15.1\times$ improvements in accuracy per power cost.
\end{abstract}

\section{Introduction}

Matrix-matrix multiplications (MMM) are prevalent in scientific computing, making General Matrix Multiply (GEMM) kernels in Basic Linear Algebra Subroutine (BLAS) highly relevant to the high-performance computing community. However, workloads have diverse numerical requirements, with ill-conditioned linear systems needing high-precision arithmetic for accurate, reproducible results~\cite{beliakov_parallel_2013,zhang_qed_1996,ellis_one-loop_2009,he_using_2000,lake_sir_1996,collange_full-speed_2014,iakymchuk_reproducible_2015,bailey_integrals_2006,bailey_high-precision_2009}, while deep neural networks exhibit resilience to arithmetic modifications and precision reduction~\cite{johnson_rethinking_2018,courbariaux_binarized_2016}.

To address IEEE-754 standard limitations, new computer formats offer different trade-offs, such as Bfloat16~\cite{abadi_tensorflow_2016,kalamkar_study_2019}, Tapered Floating-Point (TFP)\cite{Morris71}, Posit\cite{gustafson_beating_nodate}, and FP8-E4M3 and FP8alt-E5M2 formats~\cite{micikevicius_fp8_2022}. Studies compare these formats in terms of circuit area and numerical stability~\cite{jain_clarinet_2020,arunkumarm_perc_2020,tiwari_peri_2021,chaurasiya_parameterized_2018,zhang_design_2020,uguen_evaluating_2019,dedinechin:hal-01959581,buoncristiani_evaluating_2020,coding_lindstrom_2018,mallasen_quintana_percival_2021}.

However, the intermediate precision of internal arithmetic datapaths is a crucial aspect to consider in the case of GEMMs.
Because GEMMs are composed of vector-dot-products, which are arbitrary long accumulations, and because floating point addition is not transitive, the resulting calculations can significantly vary.
Several works have proposed the use of large scratchpad accumulators to address this issue~\cite{uguen_hls_2017,uguen:tel-02420901,de_dinechin_fpga-specific_2008,gustafson_beating_nodate,kulisch,ledoux_generator_2022}, though these solutions have not been widely adopted by general purpose CPUs.

Alternatively, the reconfigurable nature of FPGAs makes it possible to explore the use of workload-adaptive accumulators to introduce a controlled amount of noise in order to reduce energy costs.
However, there are currently no proposals that allow the adjustment of accumulators while measuring the impact on end-to-end workloads in terms of energy and accuracy.
Despite the extensive research in this area, the issue remains unresolved.

To address this problem, this paper presents the following contributions beyond the current state-of-the-art:

\noindent
\textbullet~~An open-source\footnote{https://github.com/Bynaryman/OSFNTC} software/hardware co-designed framework enabling intuitive intermediate precision adjustments in high-end software code, independent of the programming language, without requiring code modifications.
Our approach involves modifying and leveraging external frameworks such as PyTorch~\cite{pytorch}, NumPy~\cite{harris2020array}, OpenBLAS~\cite{xianyi_model-driven_2012,wang_augem_2013}, oc-accel~\cite{noauthor_oc-accel_2022}, OpenCAPI~\cite{stuecheli_ibm_2018}, and flopoco~\cite{flopoco1,istoan_automating_2017,DinechinPasca2011-DaT}.
This framework facilitates accuracy and energy tradeoffs by adjusting LUTs/FFs/DSPs in arithmetic datapaths and seamlessly exposes automated pipeline systolic MMM kernels to the software code (see Section~\ref{sec:framework}).

\noindent
\textbullet~~Results demonstrate substantial energy savings during validation dataset inference without compromising Top1 and Top5 scores compared to Bfloat16 and IEEE754 single-precision FMAs.
Our framework achieves energy savings of $3.3\times$ and $1.4\times$ Watt-hours for ResNet50~\cite{resnet} using ImageNet~\cite{deng_imagenet_2009}, while preserving Top1 accuracies of $82.3\%$ and $86\%$ for IEEE754-32 and BrainFloat16, respectively (see Section~\ref{sec:eval_AI}).

\noindent
\textbullet~~Our methodology reveals that double- and quad-precision arithmetic inadequately address Sea Surface Height (SSH) calculation requirements in terms of numerical correctness and reproducibility~\cite{he_using_2000,ding_data_1999}.
Using a 91-bit accumulator providing full bit-correct results and reproducibility, we achieve 5x and 27.7x more correct bits than quad- and double-precision FMAs.
This results in significant power savings, as the 91-bit accumulator generates 5.6x and 15.1x more correct bits per watt compared to quad- and double-precision FMAs, respectively (see Section~\ref{sec:eval_ssh}).


\begin{figure*}[hb!]
\vspace{-0.5cm}
  \centering
  \includegraphics[width=\linewidth, height=0.26\textheight]{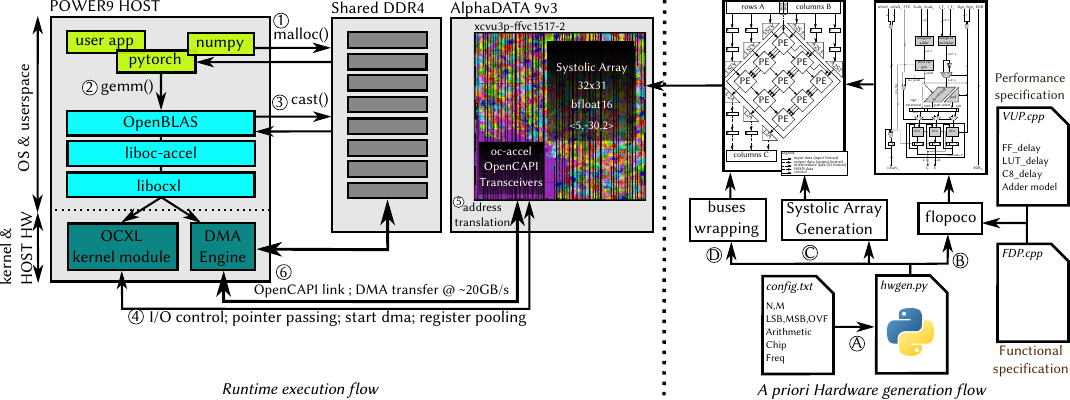}  
  \vspace{-0.3cm}
	  \caption{The 2 phases of the framework: right, the a priori Hardware generation flow, and left, the runtime execution flow.}
  	\vspace{-0.3cm}
  \label{fig:overall_framework}
\end{figure*}

\section{}
\label{sec:framework}
As depicted by Fig.~\ref{fig:overall_framework}, our framework is composed of two distinct phases, the a priori Hardware generation flow and the runtime execution flow. This Section describes both of them.

\subsection{A priori Hardware generation flow}
\vspace{-0.1cm}

\subsubsection{Functional and Performance specifications}
\label{sec:specifications}
Inspired by previous works~\cite{de_dinechin_fpga-specific_2008,uguen:tel-02420901,ledoux_generator_2022}, we design a custom dot-product operator that is agnostic to the computer format and supports variations of Posits, IEEE754, and Bfloat16.
The dot-product fuses all additions of the accumulation and performs rounding only when data exits the bottom of a systolic array.
The intermediate precision of the fixed-point accumulator used in the dot-product is a key aspect of this work, and is configurable through the length of the scratchpad delimited by the parameters MSB (Most Significant Bit) and LSB (Least Significant Bit).
Another important parameter, OVF (number of Overflow bits), helps to prevent overflows that may occur in large accumulations that do not cancel each other.
Increasing OVF by one allows for safely doubling the number of accumulations without overflow.

The automated pipeline feature of \emph{flopoco}~\cite{istoan_automating_2017} is an effective tool for efficiently exploring the wide range of the aforementioned functional specifications along with performance specifications to produce MMM kernels with the necessary basic elements (LUTs, FFs, Carry chains, DSPs) for a targeted $(chip, frequency)$ couple (see Fig.~\ref{fig:overall_framework}-\mycircle{B}).
To provide performance specifications to \emph{flopoco}, we model the Virtex Ultrascale Plus FPGA family, as our evaluation board contains an FPGA that belongs to this category.
We measure the relevant timings of the Virtex Ultrascale Plus speedgrade-2 elements and build a C++ class that inherits the \emph{flopoco} Target class.
We also refine some formulas that compute the latency of a combinational adder based on how synthesis tools map an adder.
As previous work~\cite{istoan_automating_2017} indicates, the relationship between inter-CLB (Configurable Logic Block) signal delays and their fanouts is complex, as modern FPGAs often spend more time in routing resources than in CLB resources.
To account for this, we add an ad-hoc "typicalLocalRoutingDelay" variable to the C++ class, set to $180ps$.
This variable represents the average delay for a signal to travel between two CLBs, regardless of the fanout.\looseness=-1

\subsubsection{OpenCAPI integration}
\label{sec:integration}
To integrate OpenCAPI and scale the local timing requirements strategies to the entire array and chip, we adopt a full systolic approach that avoids global data buses and control.
In this approach, data and control are registered with Flip-Flops and propagate only to adjacent processing elements (PEs), as illustrated in Fig~\ref{fig:overall_framework}-\mycircle{C}.
The resulting data flow follows a top-to-bottom and right-to-left pattern.\looseness=-1

To facilitate integration with other parts of the framework stack, we wrap the array with AXI-streaming~\cite{axi} buses, as shown in Fig.~\ref{fig:overall_framework}-\mycircle{C}.
To create the required double handshake in both master/slave directions, we add a backpressure FIFO at the bottom of the array.
The full/empty signals of the FIFO generate the valid/ready signals.

To further simplify the integration process, we utilize the oc-accel framework, which provides an abstraction layer that connects the OpenCAPI Acceleration Function Units (AFUs) interface to the more widely-used AXI-MM~\cite{axi} (see Fig.~\ref{fig:overall_framework}-\mycircle{D}).
This allows us to connect our axi-stream arrays to the AFUs via a state machine that translates between AXI-MM and AXI-streaming.
The oc-accel framework in conjunction with its own software library handles the address translation from virtual to physical pointers from user applications.
A single "make image" command in our modified oc-accel framework automates steps (\mycircle{A} to \mycircle{D}), generating a stacked hardware configuration and producing the final bitstream.

\subsection{Runtime execution flow}
\vspace{-0.1cm}
The core objective of this work is to seamlessly integrate intermediate precision adjustments from hardware into high-end software code.
We accomplish this by considering that many HPC codes utilize BLAS libraries for MMM operations.
These libraries handle GEMM function calls and dispatch them to the available hardware.
This section details the modifications made to this standard execution flow to accommodate our generated arrays and OpenCAPI.


User-level code and numerical libraries do not require any change or recompilation step to make the GEMM calls land in our numerically tailored MMM units.
The classical flow for an application is to allocate some virtual memory space for the input and output matrices and then make a call to one of the GEMM subroutines (sgemm, dgemm, zgemm, cgemm). These steps are illusatred by \mycircle{1} and \mycircle{2} in Fig.~\ref{fig:overall_framework}.
Often, such applications are distributed statically or dynamically linked with a BLAS library.
Listings~\ref{list:call_gemms}~and~\ref{list:gemm_py} show the dynamic approach to make it work with our units.
\begin{minipage}[!bt]{\columnwidth}
\begin{lstlisting}[caption={Calling twice the same user application (python) but with distinct BLAS libraries (ours and default)}, label={list:call_gemms}]
LD_LIBRARY_PATH=/opt/lib/our_openblas.lib ./gemm.py
LD_LIBRARY_PATH=/opt/lib/OpenBLAS.lib ./gemm.py
\end{lstlisting}
\vspace{-0.5cm}
\begin{lstlisting}[caption={gemm.py: a python program calling GEMMs. Our framework does not require any change in the high-level code.},label={list:gemm_py}]
import numpy as np

m,n,k = 1024, 1024, 1024
# calls dgemm
A = np.random.random((m,k))
B = np.random.random((k,n))
C = np.matmul(A, B)

# calls sgemm
A = np.random.random((m,k)).astype(np.float32)
B = np.random.random((k,n)).astype(np.float32)
C = np.matmul(A, B)
\end{lstlisting}
\vspace{-0.1cm}
\end{minipage}
%
We leverage the open-source OpenBLAS library to intercept and modify GEMM calls.
OpenBLAS has two main components: \textit{interfaces} with all GEMM function headers and \textit{backends} implementing MMMs with specific code optimizations to underlying architectures.
We add a custom backend to integrate our FPGA and designs with OpenCAPI and oc-accel.
Our modified interface and backend handle input matrix pointers, apply necessary adjustments like alignment and padding, and read the computer format used by the FPGA kernel from a configuration register.
If the format differs from the input matrices, a cast operation is performed (see \mycircle{3} in Fig~\ref{fig:overall_framework}).
This allows us to integrate our designs seamlessly with OpenBLAS and support a variety of computer arithmetics.

\section{Evaluation}
\label{sec:evaluation}

In this section, we experiment with accumulator/arithmetic combinations to showcase the trade-offs between accuracy and energy consumption in our designs for different scenarios.
We use two families of real HPC workloads with different numerical requirements to demonstrate the importance of every intermediate bit of precision in our designs.
In Section~\ref{sec:eval_ssh}, we examine a workload that is sensitive to precision, the Sea Surface Height computation~\cite{he_using_2000,ding_data_1999}.
In Section~\ref{sec:eval_AI}, we evaluate the resiliency of AI workloads in terms of Top1/Top5 accuracy when altering the sizes of internal accumulators.

\begin{figure}[t]
  \centering
	\includegraphics[width=\columnwidth]{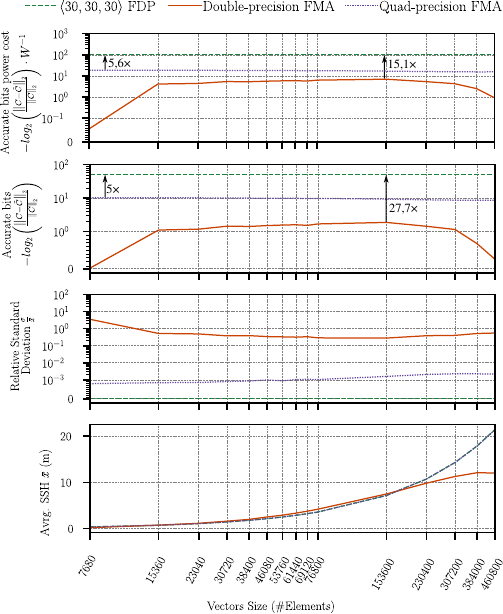}
  	\vspace{-0.5cm}
	  \caption{Sea Surface Height computation comparing IEEE-754 double-, quad- pecision FMAs and a 91-bit FDP.}
  	\vspace{-0.6cm}
  \label{fig:ssh_reproducibility}
\end{figure}

\subsection{Sea Surface Height (SSH)}
\label{sec:eval_ssh}
A relevent metric in ocean circulation model development is the Sea Surface Height (SSH) as it allows to monitor ocean current, eddies and climate changes~\cite{wunsch_satellite_1998,barrick_ocean}.
The SSH variable is a measure of the sea surface volume, which is the product of the integrated sea surface area and sea surface height.
In order to calculate the average SSH, the global sum of the sea surface volume at each model grid must be computed at each step of the computation (see Listings~\ref{list:ssh_long} and~\ref{list:ssh_lat}).
However, the data involved in the local SSH at any given grid location has an order of magnitude between $10^{10}$ and $10^{15}$ with alternating sign, while the global sum has an order of magnitude $10^{0}$.
This difference in magnitude makes the resulting average SSH impossible to interpret and reproduce using standard IEEE754-64 numbers if accumulation is made naively~\cite{he_using_2000}.
Indeed, there are several factors that can affect the order of data accumulations in a computational system, such as parallelization, out-of-order CPU execution, or in this case algorithmic rewriting to swap orders of longitude and latitude loops for instance.

There are several algorithmic and code-level techniques that can be used to reduce errors and approximate the correct result.
Self-Compensated and Double-Compensated Summations (SCS and DCS)~\cite{kahan_pracniques_1965,priest_algorithms_1991} involve estimating the round-off error at each step and subtracting it at subsequent steps.
Sorting the values in decreasing magnitude order can also be effective in this case, where values alternate signs.

An alternative is software-based extended precision emulation.
However, as prior research~\cite{bailey_high-precision_2009} indicates, this approach hampers performance.
Without dedicated circuitry, double-double precision (31 accurate significant digits) is $5\times$ slower than double precision (16 significant digits)\cite{hida_library_2007}.
This ratio increases to $25\times$ for quad-double precision and to $1000\times$ for 1000-digit arithmetic used in experimental mathematics~\cite{bailey_integrals_2006}.

\begin{minipage}[t]{\columnwidth}
\vspace{-0.4cm}
\begin{minipage}[t]{0.48\linewidth}
\begin{lstlisting}[caption={latitude first SSH calculation pseudo-code},label={list:ssh_lat}]
for i=1,64:  # latitude
for j=1,128: # longitude
    sum = sum + ssh(i,j)
end_for
end_for
print(sum) # 34.4
\end{lstlisting}
\end{minipage}%
\hfill
\begin{minipage}[t]{0.5\linewidth}
\begin{lstlisting}[caption={longitude first SSH calculation pseudo-code},label={list:ssh_long}]
for j=1,128: # longitude
for i=1,64:  # latitude
    sum = sum + ssh(i,j)
end_for
end_for
print(sum) # 0.7
\end{lstlisting}
\end{minipage}
\vspace{-0.1cm}
\end{minipage}

\begin{figure*}[tb]
  \vspace{-0.6cm}
  \centering
  \includegraphics[width=\linewidth]{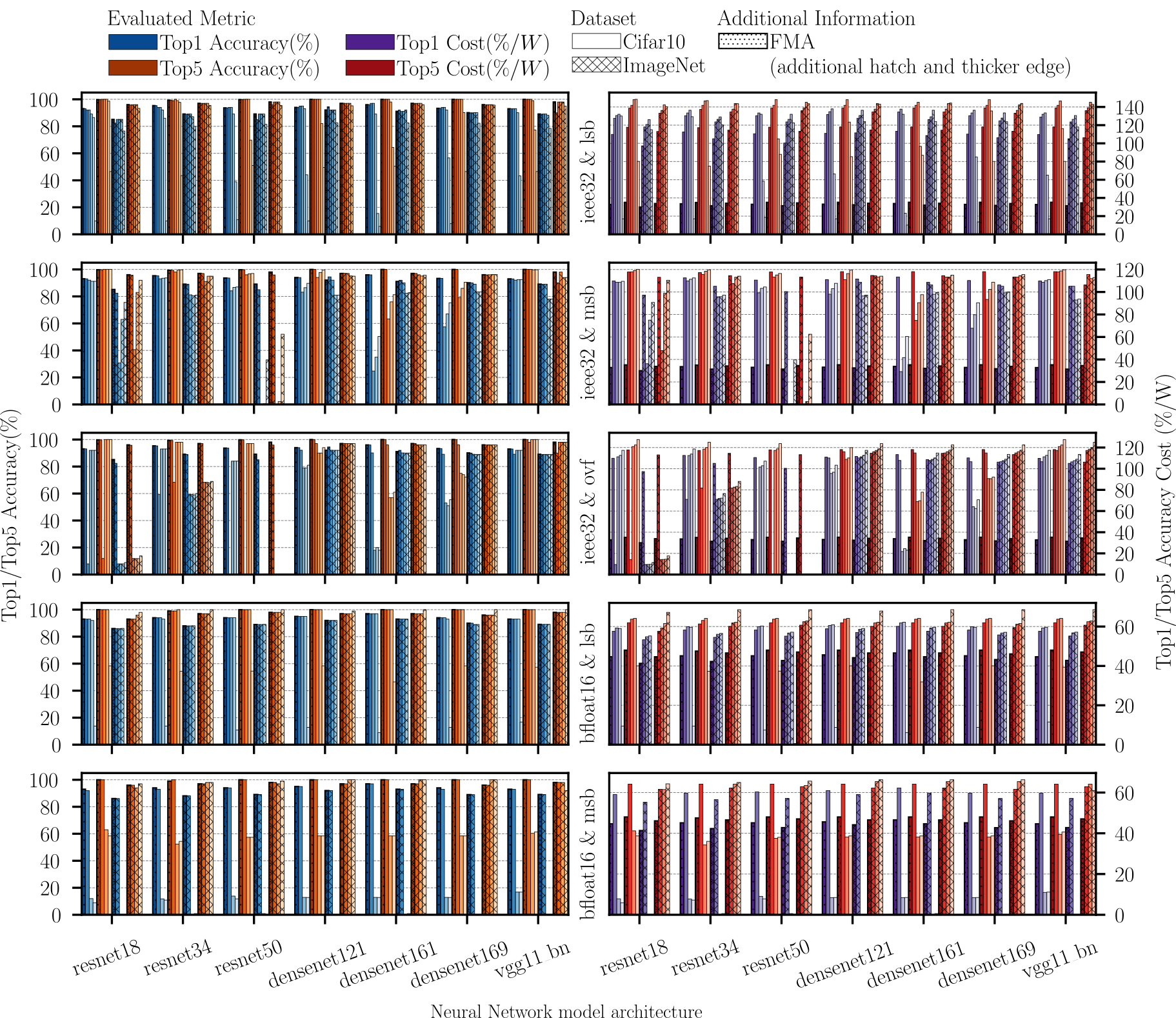}
  \vspace{-0.5cm}
	  \caption{Top1/Top5 Accuracies and Top1/Top5 Accuracy Costs for various datasets, models, computer formats, and accumulators.}
  	\vspace{-0.6cm}
  \label{fig:accuracy_and_accuracy_cost}
\end{figure*}

This study focuses on the effectiveness of hardware units, comparing our FDPs (Fused Dot Products) to double- and quad-precision FMAs found in computational systems rather than algorithmic trade-offs.
More precisely, the three compared hardware units are the IEEE-754 double-precision FMA, the IEEE-754 quad-precision FMA, and our 91-bit $\left \langle ovf:30,msb:30,lsb:30 \right \rangle$ FDP fed with IEEE754-64 words.
Figure~\ref{fig:ssh_reproducibility} presents data describing the average, the relative standard deviation (RSD), the accuracy, and the power cost per accurate bit of the SSH variable for different vector sizes.
For each grid-size, the values within the dot-products are shuffled 1000 times to observe the SSH variable spread and get a comprehensive evaluation of its reproducibility.
To ensure a fair comparison we measure the correct significant bits once the output rounded to IEEE 754 double-precision.
The results obtained with 64-bit and 128-bit FPUs exhibit decreasing reproducibility as the vector size increases.
In contrast, our 91-bit $\left \langle ovf:30,msb:30,lsb:30 \right \rangle$ FDP maintains reproducibility for all vector sizes without deviation (see the two bottom rows of Fig.~\ref{fig:ssh_reproducibility}).
Furthermore, quad-precision FPUs improve numerical quality over double-precision FPUs, but still exhibit a nearly constant RSD of $10^{-3}$, thus not providing reproducibility.
Our proposed FDP consistently exhibits 52 correct bits, which is at least 5$\times$ and 27.7$\times$ more than quad-precision and double-precision, respectively (see the second row from top of Fig.~\ref{fig:ssh_reproducibility}).
Additionally, we measure the cost of one correct bit in terms of power consumption, given by the ratio of the number of accurate significant bits to watts drawn by one unit.
Our measurements on VU3P-2 FPGA at $200MHz$ show that the units power consumption are 0.266, 0.549, and 0.491 watts for double-precision FMA, quad-precision FMA, and the 91-bit FDP, respectively.
For instance, for vectors sizes of 153600 elements, the 91-bit FDP yields 52 correct bits with a power cost of 0.491 watts, which gives a ratio of $52/0.496=104.8$, being $15.1\times$ better than the double-precision FMA ratio of $1.874/0.266=7$.
For all evaluated sizes, the 91-bit FDP yields at least $5.6\times$ and $15.1\times$ more correct bits for the same wattage as quad-precision and double-precision FMAs, respectively (see the top row of Fig.~\ref{fig:ssh_reproducibility}).
Overall, our results demonstrate that a sufficiently precise accumulator provides reproducibility and more accuracy in HPC workloads at a lower cost than double and extended precision methods.

\begin{figure*}[tb]
  \centering
  \includegraphics[width=\linewidth]{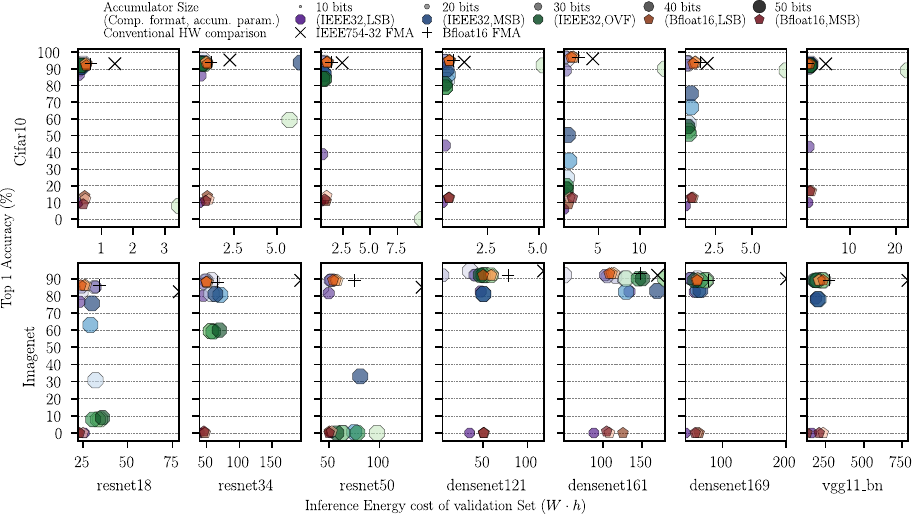}
  \vspace{-0.7cm}
	  \caption{Top1 Accuracy vs. energy cost of inferring validation datasets with various model, computer format, and accumulators.}
  	\vspace{-0.65cm}
  \label{fig:accuracy_vs_energy}
\end{figure*}

\subsection{Artificial Intelligence}
\label{sec:eval_AI}
We evaluate the accuracy and power trade-offs of low-precision accumulators across various neural network models, datasets, and computer formats.
Our focus lies on the inference portion of neural network computation, utilizing pre-trained neural networks in their original floating-point formats.

We employ Pytorch as a base framework and link it to our modified OpenBLAS.
We use popular neural network models such as ResNet18, ResNet34, ResNet50, DenseNet121, DenseNet161, DenseNet169, and VGG11 with batch normalization, and evaluate them on the CIFAR-10 and ImageNet datasets.
To measure power consumption and accuracy, we use the BrainFloat16 and IEEE-754 32-bit formats for our computations with a large variety of accumulators varying their $OVF$, $MSB$, and $LSB$ parameters, which makes a total of five categories of accumulator/arithmetic experiments.
For instance, for IEEE-754 32-bit with varying $LSB$ (first row of Fig.~\ref{fig:accuracy_and_accuracy_cost}), we evaluate the following accumulators: $\left \langle ovf:9,msb:6,lsb:-48 \right \rangle$, $\left \langle ovf:9,msb:6,lsb:-38 \right \rangle$, $\left \langle ovf:9,msb:6,lsb:-28 \right \rangle$, $\left \langle ovf:9,msb:6,lsb:-24 \right \rangle$, $\left \langle ovf:9,msb:6,lsb:-20 \right \rangle$, and $\left \langle ovf:9,msb:6,lsb:-10 \right \rangle$.

The left column of Fig.~\ref{fig:accuracy_and_accuracy_cost} shows the accuracy of the validation set of all five categories along with classic FPU-style floating-point FMA for IEEE-75 32-bit and Bfloat16.
In contrast, the right column of this Figure shows for the same configurations their accuracy cost, which we define by being the ration between the Top1 or Top5 accuracy ($\%$) and the average power drawn ($W$) by the FPGA.
Each model is represented by four groups of bars, representing from left to right the Cifar10 Top1 accuracy, Cifar10 Top5 accuracy, Imagenet Top1 accuracy, and Imagenet Top5 accuracy, respectively.
Within each group of bars, accumulator configurations are arranged from largest to smallest from left to right and are represented by progressively lighter shades of the same color.

Our experiments reveal some interesting trends.
For example, we observe that the cost of achieving one percent of Top1/Top5 accuracy is not constant across different models and datasets, suggesting that certain accumulator and arithmetic combinations perform better.
For example, the first row shows that slowly lowering $LSB$ with IEEE-754 32-bit format does not sacrifice too rapidly accuracy, but quicly exposes extrema in terms of accuracy cost.
Indeed, in the case of Resnet18 for Cifar10 and Imagenet, we can see the Top1 accuracies slowly decreasing from $93.07\%$ to $86.5\%$, and from $82.39\%$ to $76.5\%$, respectively.
However, for such cases, we averagely gain 20\% of Top1 accuracy per Watt from $LSB$ lowering between the two following accumulators that reach similar accuracies: $\left\langle ovf:9,msb:6,lsb:-48 \right\rangle$ and $\left\langle ovf:9,msb:6,lsb:-20 \right\rangle$.
This trend is observed in the majority of model, datasets, arithmetics, accumulators combinations, and is depicted by maxima bars on the right column.

In general, our experiments show that our proposed FDPs can achieve comparable or slightly better Top1/Top5 accuracy scores compared to traditional FMA, while significantly improving the accuracy per watt of power consumed.
This is clearly demonstrated in Fig.~\ref{fig:accuracy_and_accuracy_cost}, which illustrates that one of our proposed FDP configurations consistently surpasses traditional FMA in terms of Top1/Top5 Accuracy per Watt.
As an example, when applied to ResNet50 on Imagenet, IEEE754-32 and Bfloat16 FMA reach $82.3\%$ and $86\%$ Top1 accuracies, while their corresponding $\left\langle ovf:9,msb:6,lsb:-20 \right\rangle$ and $\left\langle ovf:5,msb:5,lsb:-20 \right\rangle$ FDP reach $89.1\%$ and $88.1\%$, respectively.
However, the formers respectively yield $3.3\times$ and $1.4\times$ more Top1 accuracy percentages per watt.

We also observe that some bits cannot be omitted for certain models and datasets without sacrificing accuracy.
For instance, we find that overflow bits for IEEE-754 32-bit (see third row of Fig.~\ref{fig:accuracy_and_accuracy_cost}) drastically hinder accuracies for ResNet18 and ResNet50, for all evaluated models (less than 10\% of Top1/Top5 accuracies).
This has a direct impact on the accuracy-cost ratio as overflow bits are cheap in such accumulators, resulting in poor accuracy at a low power cost gain.
However, lowering the $OVF$ parameter does not consistently lower the accuracy, as it performs correctly for some neural network models.
For example, we find that the Top1 and Top5 accuracies do not decrease in the case of DenseNet169 combined with the ImageNet dataset when lowering $OVF$, resulting in a 10\% gain in scores for the same power cost.

In addition to measuring power consumption and accuracy, we compute the energy cost in Watt-hours ($W\cdot h$) of inferring the entirety of the validation datasets.
This metric highlights the diverse profiles of various accumulators.
Fig.~\ref{fig:accuracy_vs_energy} illustrates the relationship between power consumption and accuracy for different accumulator and arithmetic combinations, revealing important trade-offs when using low precision accumulators.
The more power-efficient a combination is, the further it is located to the left of a subplot. Similarly, the more accurate a combination is, the higher it is located on the subplot.
This enables direct comparison and helps identify the best combination based on accuracy and power budgets.
For example, if $84\%$ Top1 accuracy is satisfying for Imagenet with Resnet50, the most suited arithmetic/accumulator combination is IEEE-754 32-bit/$\left\langle ovf:9,msb:6,lsb:-20 \right\rangle$ represented by a light purple hexagon as all other markers are either on the right or below.\looseness=-1

Overall, our experiments demonstrate the importance of every internal bit and the potential of using low-precision accumulators for AI workloads.
By carefully tuning the parameters of these accumulators, we are able to achieve significant power savings without sacrificing accuracy.
Furthermore, the problem can be approached from the other direction, starting with an accuracy budget and then choosing the most efficient combination for a particular dataset and model.

\section{Related Works}
\label{sec:related}

Early systolic arrays originated during the 1980s and were mostly used for convolutions~\cite{kung_systolic_1981}, linear systems~\cite{kung_why_1982} and matrix multiplications~\cite{quinton_new_nodate,kung_systolic_1978}.
Because it is not trivial to map a time-sequential algorithm into an ad-hoc space-time hardwired algorithm, previous works propose to formalize this process~\cite{sun-yuan_kung_optimal_1987}~\cite{aso_formal_1988}~\cite{isca_NavarroLV86}.
Recent AI algorithms' computational demands have rekindled interest in systolic arrays.

Genc et al.\cite{genc_gemmini_2019} introduce Gemini, an ASIC systolic array generator for RISC-V~\cite{waterman_design_2016}.
The generator, written in Chisel~\cite{bachrach_chisel_2012} HDL, supports any computer format with a Scala implementation.
However, the manuscript does not mention Chisel operators for controlling internal precision.

Chen et al.~\cite{chen_matrix-multiply_2018} design a MMM unit leveraging large accumulators.
They hardcode and optimize only one arithmetic: the posit$\left \langle 32,2 \right \rangle$ and the corresponding standardized \emph{quire} of 512 bits.
The experiments are evaluated in a POWER8~\cite{POWER8} system, with CAPI1~\cite{stuecheli_capi_2015} as link, and a VX690(28nm)~\cite{virtex-7} FPGA.
Similarly to us, they make profit of (Open)CAPI to access the entire system shared memory without the use of intermediate and non coherent representations of the data.

Likewise, Ledoux et al.~\cite{ledoux_generator_2022} propose the unique other work on generating MMM kernels that use variable precision accumulators.
They only evaluate three categories of accumulators but with various computer formats.
Unlike this work, their evaluation is limited to values within $[-1,+1]$ which does not reflect the behavior of real workloads.




%
%

\section{Conclusions}
In conclusion, our work demonstrates the potential of tailored-precision accumulators for a variety of HPC workloads, including both Artificial Intelligence and Sea Surface Height computations.
By meticulously adjusting the parameters of these accumulators, we can achieve substantial power savings without compromising accuracy or reproducibility.
Our experiments demonstrate that the ideal precision depends on the unique characteristics of each specific workload, and that intermediate precision tuning can provide a balance between power consumption and accuracy.

Ultimately, our work highlights the importance of numerically tailored accumulators for reproducibility in scientific computing applications.
Our findings offer valuable insights into the trade-offs between power efficiency and accuracy.
We are confident that our results can help guide the development of future HPC systems, and we encourage fellow researchers to investigate the potential of low-precision accumulators using our open-source framework.
\placetextbox{0.505}{0.07}{Marc Casas has been partially supported by the Grant RYC-2017-23269 funded by MCIN/AEI/10.13039/501100011033 and by ESF Investing in your future.}
\placetextbox{0.382}{0.057}{ This research was supported by grant PID2019-107255GB-C21 funded by MCIN/AEI/ 10.13039/501100011033.}
\placetextbox{0.415}{0.044}{Els autors agraeixen el suport del Departament de Recerca i Universitats de la Generalitat de Catalunya al Grup de Recerca}
\placetextbox{0.385}{0.031}{"Performance understanding, analysis, and simulation/emulation of novel architectures" (Codi: 2021 SGR 00865).}


\bibliographystyle{IEEEtranS}
\bibliography{references}

\end{document}